\documentclass[%
 reprint, amsmath, amssymb, aps, superscriptaddress, prx]{revtex4-2}
\usepackage{graphicx}% Include figure files
\usepackage{bm}% bold math
\usepackage{siunitx}
\usepackage{soul} 
\usepackage{braket}
\usepackage{setspace}
\usepackage{bm, dsfont}% bold math
\usepackage[dvipsnames]{xcolor} 
\definecolor{mycolor}{RGB}{105, 103, 175}
\usepackage[colorinlistoftodos]{todonotes}
\usepackage[colorlinks=true, linkcolor=mycolor, citecolor=mycolor, urlcolor=mycolor, breaklinks=true]{hyperref}
\usepackage{color}
\usepackage[utf8]{inputenc}
\usepackage{ulem}
\usepackage{soul}
\usepackage{epstopdf}

\begin{document}

\title{Observation of collapse and revival in a superconducting atomic frequency comb}

\author{E.S.~Redchenko}
\email{elena.redchenko@tuwien.ac.at}
\affiliation{Institute of Science and Technology Austria, Am Campus 1, 3400 Klosterneuburg, Austria}
\affiliation{Vienna Center for Quantum Science and Technology, Atominstitut, Vienna University of Technology (TU Wien), Stadionallee 2, 1020 Vienna, Austria}
\author{M.~Zens}
\affiliation{Institute for Theoretical Physics, Vienna University of Technology (TU Wien),
Wiedner Hauptstraße 8-10/136, 1040 Vienna, Austria}
\author{M.~\v{Z}emli\v{c}ka}
\affiliation{Institute of Science and Technology Austria, Am Campus 1, 3400 Klosterneuburg, Austria}
\author{M.~Peruzzo}
\affiliation{Institute of Science and Technology Austria, Am Campus 1, 3400 Klosterneuburg, Austria}
\author{F.~Hassani}
\affiliation{Institute of Science and Technology Austria, Am Campus 1, 3400 Klosterneuburg, Austria}
\author{H.S.~Dhar}
\affiliation{Department of Physics, Indian Institute of Technology, Bombay, Powai, Mumbai 400076, India}
\affiliation{Centre of Excellence in Quantum Information, Computation, Science and Technology,
Indian Institute of Technology Bombay, Mumbai 400076, India}
\author{D.O.~Krimer}
\affiliation{Institute for Theoretical Physics, Vienna University of Technology (TU Wien),
Wiedner Hauptstraße 8-10/136, 1040 Vienna, Austria}
\author{S.~Rotter}
\affiliation{Institute for Theoretical Physics, Vienna University of Technology (TU Wien),
Wiedner Hauptstraße 8-10/136, 1040 Vienna, Austria}
\author{J.M.~Fink}
\email{johannes.fink@ist.ac.at}
\affiliation{Institute of Science and Technology Austria, Am Campus 1, 3400 Klosterneuburg, Austria}
\date{\today}

\begin{abstract}

Recent advancements in superconducting circuits have enabled the experimental study of collective behavior of precisely controlled intermediate-scale ensembles of qubits. In this work, we demonstrate an atomic frequency comb formed by individual artificial atoms strongly coupled to a single resonator mode. 
We observe periodic microwave pulses that originate from a single coherent excitation dynamically interacting with the multi-qubit ensemble. We show that this revival dynamics emerges as a consequence of the constructive and periodic rephasing of the five superconducting qubits forming the vacuum Rabi split comb. In the future, similar devices could be used as a memory with \textit{in-situ} tunable storage time or as an on-chip periodic pulse generator with non-classical photon statistics.   
 
\end{abstract}

\maketitle

Collapse and revival of quantum excitations appear in many systems under different conditions \cite{mahrlein2020birth,ferreira2021collapse,eberly1980periodic,rempe1987observation,brune1996quantum,hahn1950spin,abella1966photon} and have important imaging and memory applications \cite{jung2013spin,afzelius2009multimode}. One can modify the spontaneous exponential decay of an exited system into revival dynamics using spatial and temporal interference between the atoms \cite{mahrlein2020birth} or by coupling the system to a non-Markovian bath \cite{ferreira2021collapse}. For a two-level atom interacting with a single mode of a cavity, quantum revivals were theoretically predicted \cite{eberly1980periodic} and experimentally observed \cite{rempe1987observation,brune1996quantum} within the Jaynes-Cummings model \cite{jaynes1963comparison}. Other examples of the collapse and revival dynamics include well-known spin and photon echos \cite{hahn1950spin,abella1966photon}. The conventional magnetic resonance Hahn echo sequence can lead to the train of periodic, self-stimulated revivals in a strong coupling regime \cite{debnath2020self,weichselbaumer2020echo}, while modified photon echoes using atomic frequency combs (AFC) can be used for storing photonic quantum states in atomic ensembles \cite{afzelius2009multimode}.

The revival in the AFC case corresponds to the collective retrieval of light after storage in a solid \cite{de2008solid} and proves AFCs to be a promising memory platform primarily due to their high multimode capacity \cite{ortu2022multimode}, long storage times \cite{laplane2015multiplexed,holzapfel2020optical}, and the capability of storing quantum states \cite{saglamyurek2011broadband,laplane2017multimode}. The storage efficiency, however, has to be further enhanced for quantum repeater applications \cite{briegel1998quantum}, for instance, by coupling to a cavity \cite{jobez2014cavity}. Yet, most realizations use impedance-matched cavities since frequency combs get distorted by the cavity in the strong coupling limit \cite{afzelius2010impedance}. 

A recent theoretical proposal showed how one can get around this limitation by spectrally tuning the individual atoms and adjusting their coupling values \cite{zens2021periodic}. There has been immense progress in the microscopic addressing of the atoms within the atomic array \cite{yan2023superradiant}, but such a high degree of control of individual atom parameters is currently only offered by superconducting microwave circuits. The flexibility of the circuit design supports synthesizing various few \cite{fink2008climbing,astafiev2010resonance} and many-body interaction Hamiltonians \cite{puertas2019tunable,zhang2022synthesizing,mehta2023down}. This ability to tune the coupling strength is vital for the generation of multi-qubit entanglement in a quantum network \cite{zhong2021deterministic}. Moreover, the rapid development of superconducting circuits over the last several decades has now allowed precise control over intermediate-sized systems \cite{wang2020controllable,yang2020probing}. 

\begin{figure}[t]
\centering
\includegraphics[width=.95\columnwidth]{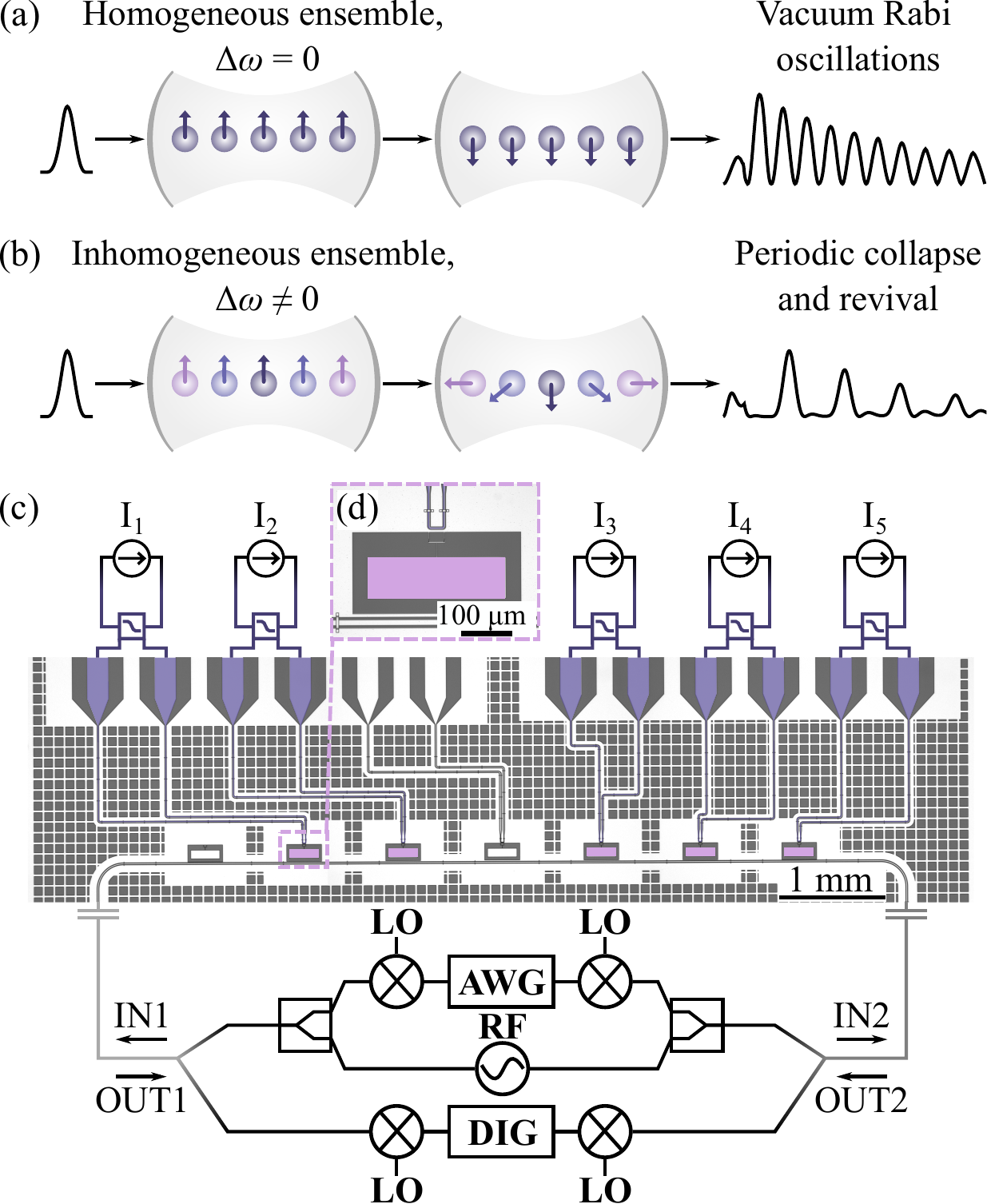}
\caption{(a) Schematic showing collective vacuum Rabi oscillations caused by a short excitation pulse in a resonant and homogeneous qubit ensemble, $\Delta\omega=0$. (b) Schematic showing periodic pulse revivals caused by a short excitation pulse in an inhomogeneous qubit ensemble, $\Delta\omega\neq0$. (c) Optical microscope image and simplified experimental setup. Seven transmon qubits are capacitively coupled to a coplanar waveguide resonator, and six qubits have a symmetric local flux bias line highlighted in purple. Qubits used in a sAFC are highlighted in pink. Each comb qubit is connected to a current source. The system gets continuous input from an RF source and pulsed input from the upconverted arbitrary waveform generator (AWG) signal. Analog downconversion and digitization (DIG) are used to back out the scattering parameters of the device cooled to 10\,mK. (d) Enlarged view of qubit and local symmetric flux bias line inductively coupled to the qubit SQUID.} 
\label{fig:pulse_1}
\end{figure}

In this Letter, we present the first superconducting AFC (sAFC) based on individual artificial atoms. Within this versatile system, superconducting transmon qubits can be individually tuned with high precision changing the inhomogeneity of the sAFC to produce cavity state revivals at well-defined and tunable times. The sAFC is realized on-chip with five transmon qubits capacitively coupled to the microwave resonator. The central qubit is set in resonance with the cavity $\omega = \omega_{\text{c}}$ while others are $\pm\Delta\omega$ and $\pm2\Delta\omega$ detuned. In contrast to a resonant ensemble, shown in Fig. \ref{fig:pulse_1}(a), where the collective vacuum Rabi oscillations are observed \cite{kaluzny1983observation,fink2009dressed}, in such an inhomogeneous ensemble (Fig. \ref{fig:pulse_1}(b)), collectively excited qubits arranged in a sAFC rotate over the Bloch sphere at different speeds. The subensembles of qubits constructively rephase at regular intervals, which leads to a periodic photon emission corresponding to the collective transfer of excitations from the qubit ensemble to the resonator \cite{dhar2018variational}.

%%%%%%%%%%%%%%%%%%%%%%%%%%%%%%%%%%%%%%%%%%%%%%%%%%%%%%%%%%%
%%%%%%%%%%%%%%%%%%%%%%%%%%%%%%%%%%%%%%%%%%%%%%%%%%%%%%%%%%%%

The studied system is described by the driven Tavis-Cummings Hamiltonian \cite{tavis1968exact}:

\begin{equation}
\hat{H}_{\text{sys}} = \hat{H}_{\text{TC}} + \hat{H}_{\text{drive}},\label{eq:H_sys}
\end{equation}
where
\begin{align}
\begin{split}\label{eq:H_TC}
    \hat{H}_{\text{TC}}={}&\hbar\omega_{\text{c}}\hat{a}^{\dagger}\hat{a}+\frac{\hbar}{2}\sum_{k=1}^N{\omega_{k}\hat{\sigma}_k^z}\\
    &+\hbar\sum_{k=1}^Ng_k(\hat{\sigma}_k^-\hat{a}^{\dagger}+\hat{\sigma}_k^+\hat{a}),
\end{split}\\
\hat{H}_{\text{drive}} = {}& i(\eta(t)\hat{a}^{\dagger}e^{-i\omega_{\text{d}} t}+\eta^*(t)\hat{a}e^{i\omega_{\text{d}} t}). \label{eq:H_drive}\, 
\end{align}

Here $g_k$ are the qubits' coupling strengths to the resonator, $\omega_k$ are the qubits' frequencies, $\eta$ is the drive amplitude, and $\omega_{\text{d}}$ is the excitation pulse frequency. 

We fabricate the sample with seven transmon qubits capacitively coupled to the $\lambda$-mode of a half-wavelength coplanar waveguide resonator (CPW) shown in Fig. \ref{fig:pulse_1}(c). For the distance between qubits of $600$\,$\mu$m, the simulated direct capacitive coupling between them $J_{\text{C}}/(2\pi)\approx1$\,kHz is negligible. We also don't observe any direct inductive coupling between qubits. Qubits used to form a sAFC are highlighted in pink. Six transmons have individual on-chip symmetric DC-bias lines inductively coupled to the SQUIDs as demonstrated in Fig. \ref{fig:pulse_1}(d). 

The $\lambda$-mode of the CPW has a resonance at $\omega_{\text{c}}/(2\pi) = 5.878$\,GHz. Two ports of the resonator are coupled with $\kappa_{\text{e}1(2)}/(2\pi)=0.42 (51)$\,MHz. The internal loss of the resonator $\kappa_{\text{i}}/(2\pi) \approx 3$\,kHz is negligible compare to the full resonator bandwidth $\kappa_{\text{load}}/(2\pi) \approx 0.93$\,MHz. The average qubit coupling strength to the resonator is $g/(2\pi)\sim30$MHz, and qubits decoherence parameter $\gamma/(2\pi)<500$kHz were extracted from vacuum Rabi splitting measurements. See full sample characterization in the Supplemental Material \cite{smpulserev}.

%%%%%%%%%%%%%%%%%%%%%%%%%%%%%%%%%%%%%%%%%%%%%%%%%%%%%%%%%%%
\begin{figure}[t]
\centering
\includegraphics[width=\columnwidth]{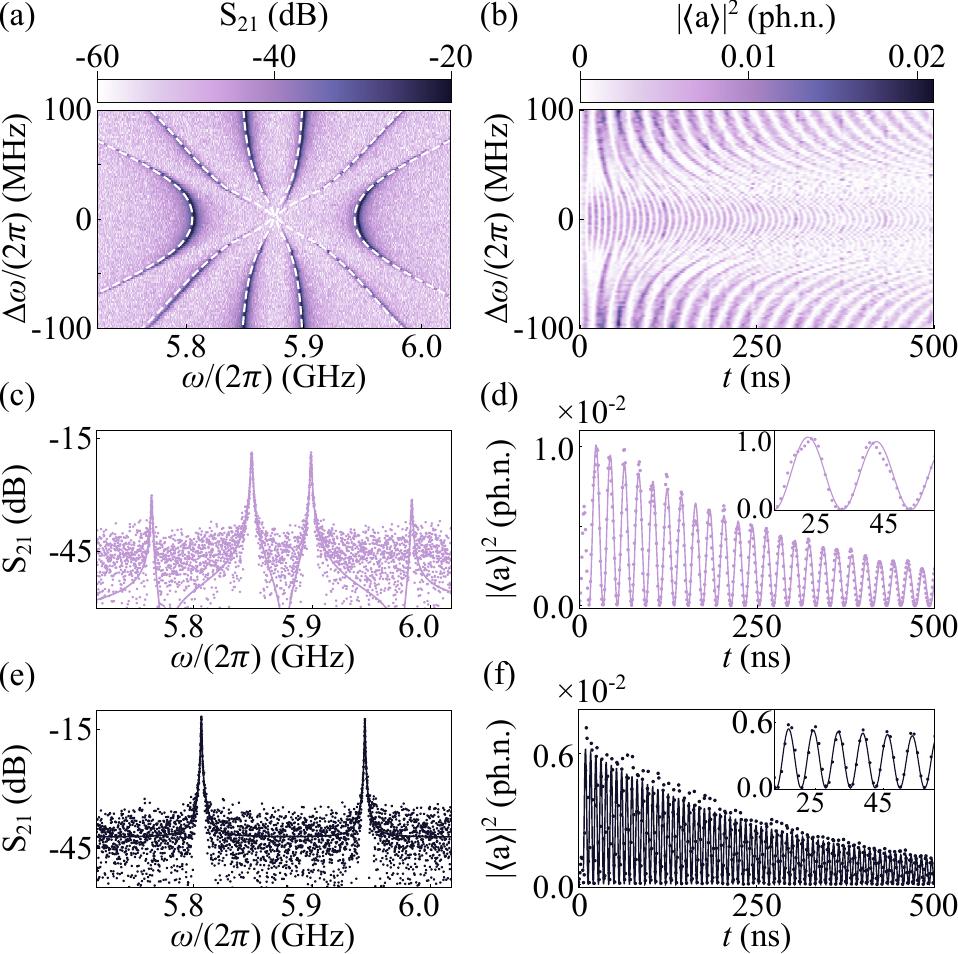}
\caption{ (a) Measured resonator transmission spectrum of a 5 qubit ensemble as a function of probe frequency $\omega$ and comb spacing $\Delta\omega$. White dashed lines are the calculated eigenvalues. (b) Squared absolute value of the transmission amplitude $|\langle a\rangle|^2$ measured as a function of time $t$ and comb spacing $\Delta\omega$. (c,e) Transmission spectrum of a 5 qubit ensemble as a function of probe frequency $\omega$ for comb spacing $\Delta\omega/(2\pi)$ of $100$\,MHz and $0$\,MHz respectively. (d,f) Single qubit ($\Delta\omega/(2\pi)=100$\,MHz) and collective ($\Delta\omega/(2\pi)=0$\,MHz) vacuum Rabi oscillations.} 
\label{fig:pulse_2}
\end{figure}
%%%%%%%%%%%%%%%%%%%%%%%%%%%%%%%%%%%%%%%%%%%%%%%%%%%%%%%%%%%%

%%%%%%%%%%%%%%%%%%%%%%%%%%%%%%%%%%%%%%%%%%%%%%%%%%%%%%%%%%%
\begin{figure*}[t!]
\centering
\includegraphics[width=2\columnwidth]{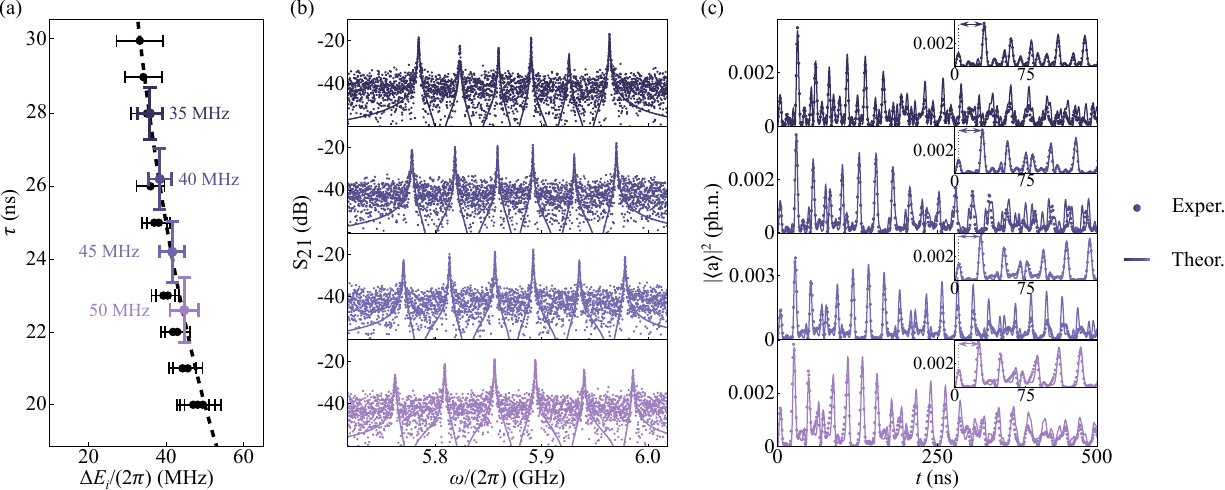}
\caption{(a) The revival time $\tau$ as a function of comb spacing $\Delta\omega$ extracted from Fig. \ref{fig:pulse_2}(b). (b) Transmission spectra of a 5 qubit ensemble measured as a function of probe frequency $\omega$ for comb spacing $\Delta\omega/(2\pi) = 35, 40,45,50$\,MHz. Solid lines are fitted with Eq.~\ref{eq:S_21_6}. (c) Squared absolute value of the transmission amplitude measured as a function of time $t$ for same detuning frequencies as shown in (b). Solid lines are numerical solutions of Eq.~\ref{eq:Lindblad}. Insets demonstrate transmitted excitation pulse and approximately five first revivals, where the first storage time is indicated with an arrow.} 
\label{fig:pulse_3}
\end{figure*}
%%%%%%%%%%%%%%%%%%%%%%%%%%%%%%%%%%%%%%%%%%%%%%%%%%%%%%%%%%%%

Symmetric DC-bias lines used in this device allow tuning the qubit frequency up to two flux quanta without introducing a significant flux cross-talk between qubits. Additional global flux offset can be created using the bias coil mounted on top of the sample box. The full flux quanta matrix and the calibration procedure can be found in the Supplemental Material \cite{smpulserev}. We demonstrate full frequency control by measuring the transmission amplitude as a function of measurement frequency/time and comb spacing $\Delta\omega/(2\pi)$, as shown in Fig. \ref{fig:pulse_2}(a,b).

In the frequency domain, we measure the transmission spectrum of a qubit ensemble versus probe frequency $\omega$ and comb spacing $\Delta\omega$, as shown in Fig. \ref{fig:pulse_2}(a) where white dashed lines show the numerically calculated eigenvalues $E_i$ of the Tavis-Cummings Hamiltonian (Eq.~\ref{eq:H_TC}). For high comb spacings $|\Delta\omega/(2\pi)|>2g/(2\pi)$, transmission is dominated by the vacuum Rabi splitting peaks of the central qubit, shown in Fig. \ref{fig:pulse_2}(c). Approaching $\Delta\omega/(2\pi)=0$, these dressed states then merge into two peaks that correspond to collective vacuum Rabi splitting with the collective coupling frequency $G/(2\pi) = \sqrt{\sum_{i=k}^5g_k^2} = 68.95 $\,MHz, shown in Fig. \ref{fig:pulse_2}(e).

In the time domain measurements, shown in Fig. \ref{fig:pulse_2}(b), the system is excited by the short ($6$\,ns) pulse which is generated using upconversion of the output signal from the AWG. The pulse amplitude $\eta\sim5\kappa_{\text{load}}$ in this measurement corresponds to the input power of $25$ photons in the case of the empty cavity with $\kappa_{\text{load}}$ bandwidth. Transmission amplitude in Volts is then recorded by the DIG at high ($250$\,MHz) intermediate frequency. For high comb spacings $|\Delta\omega/(2\pi)|>2g/(2\pi)$, we observe vacuum Rabi oscillations of the central qubit with frequency $\omega_{\text{Rabi}} = 49.6 $\,MHz $\approx 2g/(2\pi)$, as shown in Fig. \ref{fig:pulse_2}(d) where the solid line is the data fit with damped sinusoidal function, $f(t) = e^{-\Gamma t}(\sin{\Omega t}+1)$. For close to zero detunings, we see the expected collective vacuum Rabi oscillations, shown in Fig. \ref{fig:pulse_2}(f). Now, the oscillation frequency extracted from the damped sinusoidal fit $\omega_{\text{colRabi}} = 137.4 $\,MHz is close to the collective vacuum Rabi splitting $2 G/(2\pi) = 137.9 $\,MHz. Moreover, in both cases of single and collective vacuum Rabi oscillations, we find that decay is dominated by the resonator coupling $\Gamma \approx 1.1$\,MHz $\sim \kappa_{\text{load}}$.

For the intermediate comb spacings $g\lesssim\Delta\omega\lesssim 2g$, we observe six bright dressed states $E_i$ of the 5-qubit ensemble coupled to the resonator in the frequency domain, see Fig. \ref{fig:pulse_2}(a). In the time domain, when the short pulse is sent to the sAFC, the system exhibits a pulsed collapse and revival dynamics. In Fig. \ref{fig:pulse_3}(a), black dots demonstrate the extracted revival time $\tau$ from Fig. \ref{fig:pulse_2}(b) defined as the hold time between the transmitted pulse and the first revival as a function of the frequency difference between the respective neighboring bright states $\Delta E_i$ ($i=1,...,6$) obtained from Fig. \ref{fig:pulse_2}(a). The revival time decay with the increase of the comb spacing and the dashed line shows the expected $\tau = 2\pi/\Delta E_i$ trend \cite{dhar2018variational,zens2021periodic}. The error bars indicate the standard deviation from the mean value of $\Delta E_i$. High error values (up to $\sim6$\,MHz for small combs) come from the spectral distortion of the sAFC due to strong coupling to the resonator \cite{zens2021periodic} while the positioning of qubits in the comb is implemented with up to $\sim1$\,MHz precision. 

Colored dots in Fig. \ref{fig:pulse_3}(a) indicate the storage time for the comb spacing of $\Delta\omega/(2\pi) = 35$, $40$, $45$, and $50$\,MHz measured separately at $\eta\sim2.5\kappa_{\text{load}}$ drive amplitude. We first measure the system in the frequency domain to identify qubits positions precisely. The transmission spectra, shown in Fig. \ref{fig:pulse_3}(b), are fitted using the following function 
\begin{equation}\label{eq:S_21_6}
S(\omega) = \frac{A}{\kappa_{\text{load}}+i(\omega_{\text{c}} - \omega)+\sum_{k=1}^5\frac{g_k^2}{\gamma_k^2+i(\omega_k-\omega)}}.
\end{equation}

The fit relies on the qubits parameters extracted from the vacuum Rabi splitting measurements ($g_k$ and $\gamma_k$). The qubits' spectral positions $\omega_k$ and the dimensionless amplitude $A$ given by attenuation and amplification in the setup are kept as free parameters. 

The time-evolution of the system can be theoretically predicted by solving the Lindblad master equation \cite{dhar2018variational}:

\begin{equation}\label{eq:Lindblad}
\Dot{\rho} = - [\hat{H}_{\text{sys}},\rho]+\kappa_{\text{load}}\mathcal{L}_{\hat{a}}[\rho]+\sum_k\gamma_k\mathcal{L}_{\sigma_k}[\rho],
\end{equation}
where $\mathcal{L}_{\hat{a}} = \hat{a}$ is the Lindblad operator corresponding to the resonator losses and the qubit decay is accounted for by $\mathcal{L}_{\sigma_k} = \sigma_k^-$. 

We numerically solve Eq.~\ref{eq:Lindblad} in Python using QuTip \cite{johansson2012qutip} for the extracted qubit positions and find the behavior of the coherent cavity response $|\langle a\rangle|^2$. Time traces shown in Fig.~\ref{fig:pulse_3}(c) were scaled to the numerically predicted value. We see excellent agreement between the theoretically predicted and experimentally observed microwave revival dynamics. The revival time for these comb spacings was extracted from the first $6$ peaks excluding one overlapping with the beat mode on top of the revivals which is caused by the slight difference in the coupling coefficients $g_k$. The error of $0.8$\,ns is below the $\sim2$\,ns time resolution in our system. The noise between revivals in Fig.~\ref{fig:pulse_3}(c) comes from the finite ensemble size where the system is found in a mix between vacuum Rabi oscillations and full constructive/destructive rephasings. The decay of the revival dynamics is dominated by the resonator coupling $\kappa_{\text{load}}$.

%%%%%%%%%%%%%%%%%%%%%%%%%%%%%%%%%%%%%%%%%%%%%%%%%%%%%%%%%%%
\begin{figure}[t]
\centering
\includegraphics[width=\columnwidth]{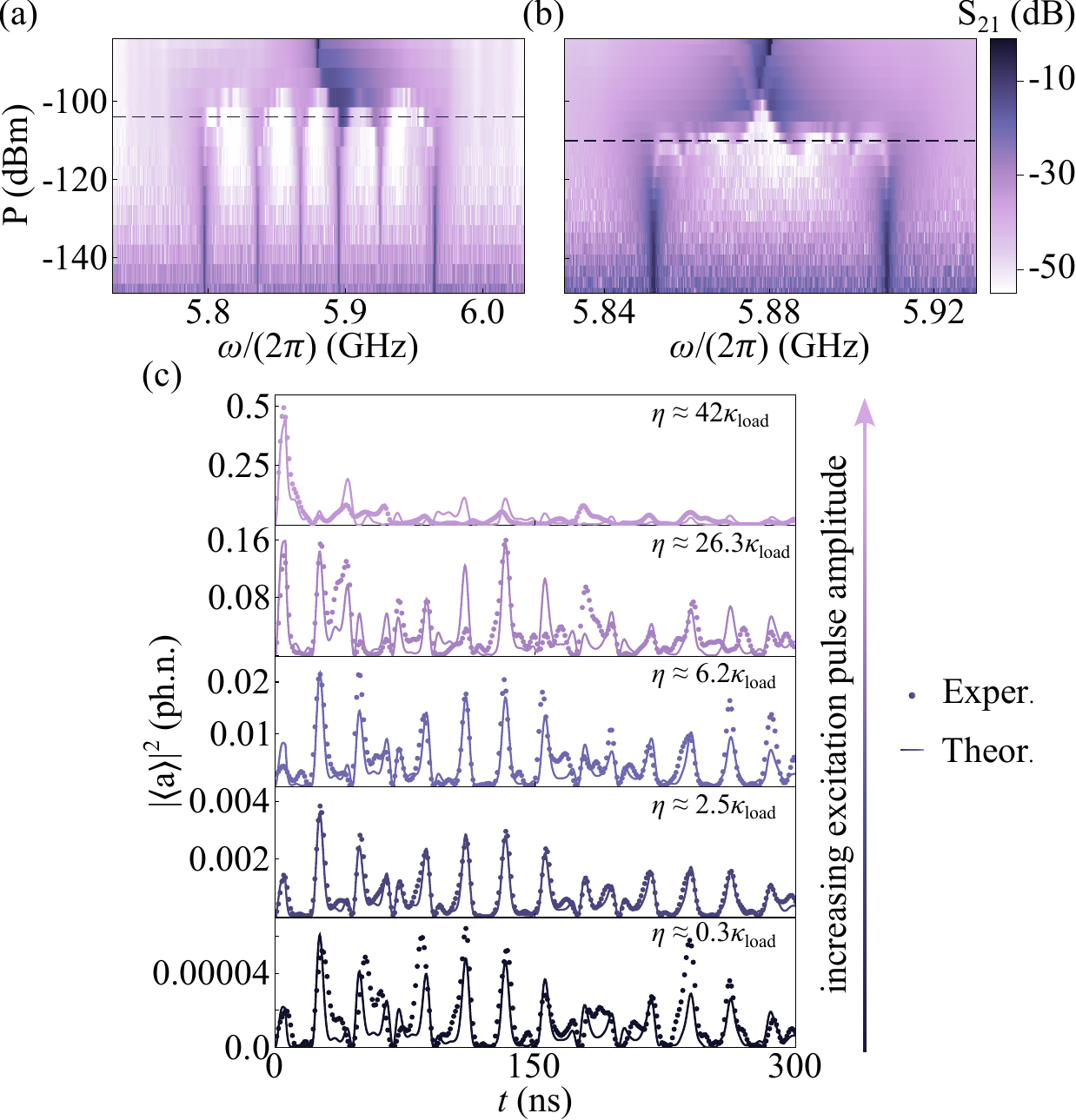}
\caption{Transmission spectra of a 5 qubit ensemble with a comb spacing $\Delta\omega/(2\pi) = 30$\,MHz (a) and a single qubit vacuum Rabi splitting (b) measured as a function of probe frequency $\omega$ and probe power P. (c) Squared absolute value of the transmission amplitude measured as a function of time $t$ for different drive amplitudes $\eta$ for comb spacing $\Delta\omega/(2\pi) = 50$\,MHz.} 
\label{fig:pulse_4}
\end{figure}
%%%%%%%%%%%%%%%%%%%%%%%%%%%%%%%%%%%%%%%%%%%%%%%%%%%%%%%%%%%%

Another interesting property of the sAFCs is the enhanced power handling. Power dependence measurements of the sAFC and a single resonant qubit in a cavity reveal that the dressed states of the qubit ensemble get saturated at least at $\sim6$\,dB higher probe power than the dressed states of a single qubit, which is indicated with black dashed lines in Fig.~\ref{fig:pulse_4}(a,b). We also observe the pulse revival dynamic to be preserved for a wide range of input powers. In fact, sending excitation pulse with an amplitude scaling from $\eta\approx 0.3\kappa_{\text{load}}$ to $\eta\approx 6.2\kappa_{\text{load}}$, corresponding to $0.1$ to a $38$ input photon power, doesn't break revivals. We observe the distortion of the pulse revivals at drive amplitude around $\eta\approx 26.3\kappa_{\text{load}}$ or $695$ photons, while the clear destruction of revival dynamics happens at $\eta\approx 42\kappa_{\text{load}}$ or $1764$ photons. The theoretically predicted coherent cavity response is shown with the solid line in Fig.~\ref{fig:pulse_4}(c). 

Overall, signatures of the constructive rephasing in the qubit ensemble are observable in a $\sim30$\,dB range of input powers. Such a high operation band can be explained by two characteristics of our system. First of all, the sAFC acts as a filter for the excitation pulse. According to our estimations based on the integrated overlap of the comb and pulse envelopes in the frequency domain only $\sim1/25$ of the excitation pulse power is absorbed by the qubit ensemble, see Supplemental Material \cite{smpulserev}. Secondly, as we deal with a multiqubit system, which has a higher cumulative absorption, it can handle higher input photon numbers before getting saturated (5 photons, in our case). Therefore, an increase in the number of qubits in the ensemble not only helps with the pulse definition but also further expands the operating power range.

In conclusion, we have studied the collapse and revival dynamics of a microwave pulse in an inhomogeneous superconducting qubit ensemble that forms the first sAFC made of individual emitters. The revival time of the observed tunable periodic pulses is controlled \textit{in-situ} by variation of the comb spacing $\Delta\omega$ through a change in the bias settings. For small comb spacings, our system with its increased storage time operates as a coherent microwave memory. Remarkably, in contrast to the realization with superconducting multiresonators \citep{bao2021demand}, it has potential applications in quantum memory protocols \cite{zens2021periodic}. When the comb spacing is large, the high repetition rate of the revivals together with the enhanced power handling enables this on-chip integrated device to be used as a periodic signal generator that outputs time-constrained pulses at regular intervals. Based on their quantum nature, sAFCs have a prospect to act as periodic non-classical microwave sources for timing in quantum networks \cite{kimble2008quantum}. Combining the demonstrated sAFC with fast time-domain control of the qubit frequencies, excitations, and tomography, one can utilize these fast interactions within sAFC for the preparation and distribution of quantum states in qubit ensembles rather than individual atoms \cite{ferreira2022deterministic,kannan2023on}. Therefore, the achieved high degree of control over multi-qubit ensembles not only allows exploring a new regime of light-matter interaction but also represents another step toward new applications in future superconducting quantum computing hardware.

The authors thank G. Arnold and R. Sahu for the discussions, L. Drmic, P. Zielinski, and R. Sett for software development, the MIBA workshop and the ISTA nanofabrication facility for technical support, and VTT Technical Research Centre of Finland for providing us TWPAs for follow-up measurements. This work was supported by the Austrian Science Fund (FWF) through BeyondC (F7105) and IST Austria. E.S.R. is the recipient of a DOC fellowship of the Austrian Academy of Sciences at IST Austria. J.M.F. and M.\v{Z}. acknowledge support from the European Research Council under grant agreement No 758053 (ERC StG QUNNECT) and a NOMIS foundation research grant. 

\renewcommand*{\bibfont}{\footnotesize}

\onecolumngrid
\newpage
\normalsize
\appendix
\begin{center}
\textbf{\large{Supplemental Material for: Observation of collapse and revival in a superconducting atomic frequency comb}}
\end{center}

\section*{Sample characterization}
\begin{figure}[h]
\centering
\includegraphics[width=0.53\columnwidth]{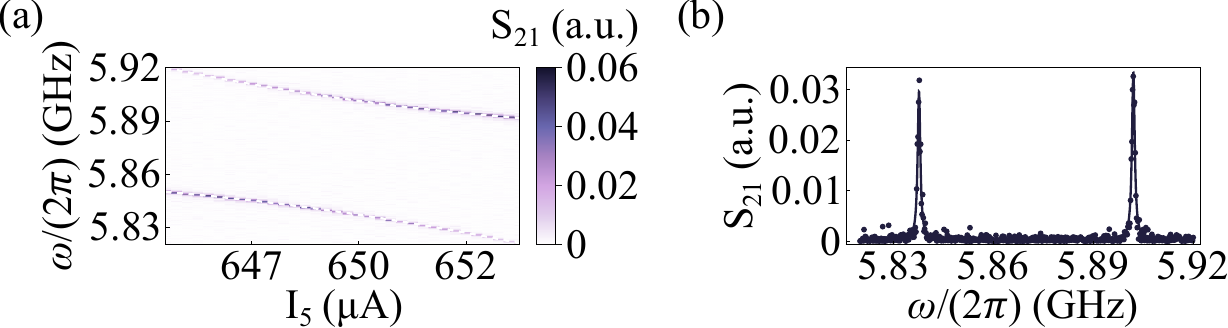}
\caption{(a) Transmission spectra measured as a function of frequency and current applied to Qubit 5 to vacuum Rabi splitting, $\omega_{5}\sim\omega_{\text{c}}$. (b) Transmission spectrum as a function of frequency at $\omega_{01}= \omega_{\text{res}}$ fitted with double Lorentzian function Eq. \ref{eq:Lor}.} \label{Sfig:pulse_1}
\end{figure}

We have designed the sample such that the coupling between each qubit and the resonator is close to $g/(2\pi)\approx30$\,MHz. To experimentally determine the coupling strengths $g_k$ of each qubit, we first park all other qubits at around their maximum frequencies. Then we tune the qubit across the resonator and measure the transmission spectrum, shown in Fig. \ref{Sfig:pulse_1}(a). We observe energy level splitting $(E_+-E_-)/\hbar = (\omega_k+\omega_{\text{c}})\pm\sqrt{(\omega_k+\omega_{\text{c}})^2+4g_k^2}/2$, where at degeneracy ($\omega_k=\omega_{\text{c}}$) with the minimal level splitting the frequency difference is given by the vacuum Rabi splitting $(E_+-E_-)/\hbar =  2g_k$. The measured coupling strengths $g_k$, see Table \ref{tab:pr1}, are in good agreement with the designed value.

%%%%%%%%%%%%%%%%%%%%%%%%%%%%%%%%%%%%%%%%%%%%%%%%%%%%%%%%%%%
%%%%%%%%%%%%%%%%%%%%%%%%%%%%%%%%%%%%%%%%%%%%%%%%%%%%%%%%%%%%

To extract the decoherence rate we fit the measured vacuum Rabi splitting, shown in Fig. \ref{Sfig:pulse_1}(b), with the double Lorentzian function
\begin{eqnarray}
\label{eq:Lor}
&&|S(\omega)|^2 = \frac{A_1}{1+(\frac{\omega-(\omega_{\text{c}}-g_k)}{\kappa_k/2})^2}+\frac{A_2}{1+(\frac{\omega-(\omega_{\text{c}}+g_k)}{\kappa_k/2})^2},
\end{eqnarray}
where $A_{1(2)}$ are dimensionless amplitudes given by attenuation and amplification in the setup, at the degeneracy point $A_1 \approx A_2$. The bandwidth of each peak $\kappa_k = (\kappa_{\text{load}}+\gamma_k)/2$ depends on the loaded cavity linewidth and the qubit decoherence rate.

\begin{table}[ht]
\begin{center}
    \begin{tabular}{ | c || c | c | c | c | c | c | c |}
    \hline
    Qubit number $k$ & 1 & 2 & 3 & 4 & 5 & 6 & 7\\ \hline\hline
    $g_k/(2\pi)$\,MHz & 28.07 & 30.96 & 32.27 & 30.82 & 32.13 & 30.54& 28.24\\ \hline
    $\gamma_k/(2\pi)$\,kHz & $<$10 & 414 & 287 & 470 & 350 & 290 & 33\\ \hline
    \end{tabular}
    \caption{\label{tab:pr1}Measured coupling strengths $g_k$ and the ectracted decoherence rates $\gamma_k$.}
\end{center}
\end{table}

\section*{Frequency comb preparation}
\begin{figure}[h]
\centering
\includegraphics[width=0.5\columnwidth]{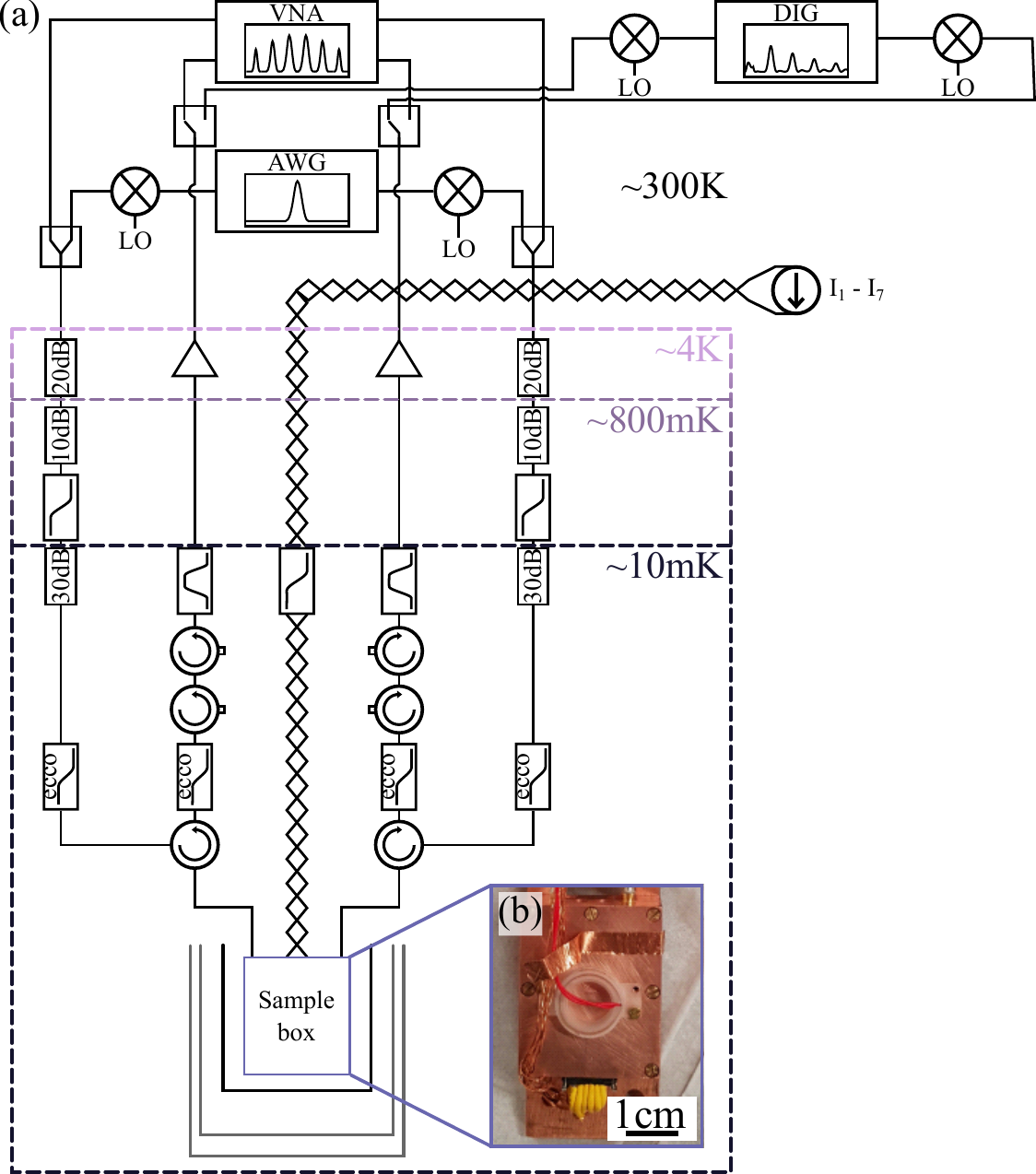}
\caption{(a) Extended schematic drawing of the measurement setup. (b) Photograph of a sample box with the bias coil mounted on top.} 
\label{Sfig:pulse_2}
\end{figure}

We start the preparation of inhomogeneous ensembles by investigating flux-cross talk and offsets. Although symmetric local flux control lines of each qubit are designed to mitigate the crosstalk between qubits and noncorresponding flux control lines, the crosstalk is still not negligible due to residual parasitic coupling in the DC wiring and on chip. Moreover, we use a bias coil mounted on a sample box, as shown in Fig. \ref{Sfig:pulse_2}(b) which tunes all qubits similarly and creates a global flux offset.

%%%%%%%%%%%%%%%%%%%%%%%%%%%%%%%%%%%%%%%%%%%%%%%%%%%%%%%%%%%
%%%%%%%%%%%%%%%%%%%%%%%%%%%%%%%%%%%%%%%%%%%%%%%%%%%%%%%%%%%%

To get full frequency control of the comb, we calibrate the system by measuring two-tone spectroscopy of qubits ensemble as a function of flux applied to each tuning channel including bias coil and fitting each measured spectral line with the following function 
\begin{eqnarray}
\label{eq:freq_off}
&&\omega_{01,k,j} \approx \omega_{01,\text{max},k} \sqrt{|\cos{(\pi(\Phi_j/\Phi_{j,k})-\pi\phi_{\text{off},k})}|},
\end{eqnarray}
where $\omega_{01,\text{max},k}$ is a maximum qubit frequency, $\Phi_j$ is a flux applied to the bias channel $j$ in A normalized to the flux quanta of a qubit in this channel $\Phi_{j,k}$, and $\phi_{\text{off}}$ is a dimensionless flux offset.

Thus, we reconstruct the full cross-talk matrix
\begin{eqnarray}
\label{eq:cr_t_m}
&&M = \frac{1}{(\Phi_{jk})_{7\times7}} = \begin{pmatrix}
-7 & 450 & -12 & -22 & -10 & 0.2 & 0.16\\
-0.16 & 2.5 & 439 & 27 & 2.5 & -0.3 & 0.1\\
-3.3 & -0.8 & -3.1 & 427 & -12 & 2 & 0.14\\
-0.33 & -0.13 & -0.13 & -4 & 421 & 6.6 & 0.9\\
0.33 & 0.05 & 0.05 & 1 & 12 & 441 & -3.6\\
0.33 & 0.02 & 0.02 & 0.02 & 2.5 & -14.9 & 437\\
12820 & 23809 & -23696 & 25974 & 26455 & 26247 & 21505\\
\end{pmatrix},
\end{eqnarray}
where $\Phi_{jk}$ is a flux quanta in A, $j$ is the number of the tuning line (6 on-chip lines $I_1 - I_6$, and 1 bias coil $I_7$), $k$ is the qubit number counting from left to right. We also find the flux offset vector
\begin{eqnarray}
\label{eq:off_t_m}
&& \phi_{off}^{\text{T}} = \begin{pmatrix}
-0.26 & 0.34 & -0.27 & -0.31 & -0.27 & 0.3 & 0.1\\
\end{pmatrix}.
\end{eqnarray}

\begin{figure}[h]
\centering
\includegraphics[width=0.55\columnwidth]{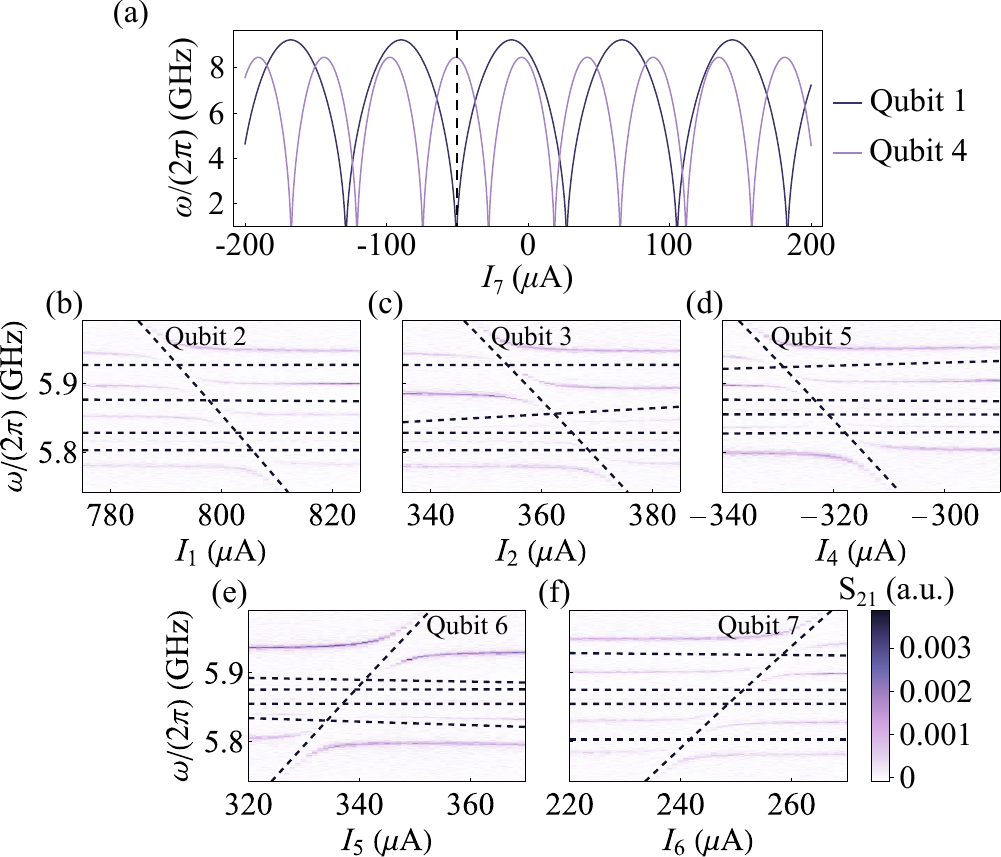}
\caption{(a) Reconstructed frequency spectra (Eq. \ref{eq:freq_off}) of Qubit 1 and Qubit 4 as a function of flux applied with the bias coil. The black dashed line shows the chosen bias point at which frequency combs will be created. (b-f) Transmission spectra of a 5 qubit ensemble as a function of frequency and flux applied with the local DC-bias lines near the frequency comb bias settings.} 
\label{Sfig:pulse_3}
\end{figure}

Now we can identify approximate qubit positions at the given bias settings and start the comb preparation. We only use five qubits in the frequency comb to observe the pulse revival dynamics, the other two qubits are tuned far away ($>2$GHz) from bare resonator frequency with the bias coil. That helps to reduce the size of the effective cross-talk matrix. We choose the coil bias point manually from the plot shown in Fig. \ref{Sfig:pulse_3}(a) which exhibits the frequency behavior of Qubit 1 and Qubit 4 as a function of flux applied to the coil. We take into account that the bias coil generates global flux which results in the new flux offsets for the qubit in the comb.
Furthermore, we see that flux offsets drift in time. Hence, to get more accurate values, we tune comb qubits closer to the bare resonator frequency $\omega_{\text{c}}$, and then sweep current applied to each local line and fit the qubit frequency using the full cross-talk matrix and adjusted the flux offsets, as shown in Fig. \ref{Sfig:pulse_3}(b-f).

%%%%%%%%%%%%%%%%%%%%%%%%%%%%%%%%%%%%%%%%%%%%%%%%%%%%%%%%%%%
%%%%%%%%%%%%%%%%%%%%%%%%%%%%%%%%%%%%%%%%%%%%%%%%%%%%%%%%%%%%

These final adjustments help us to prepare the ensemble with the desired comb spacing $\Delta\omega$, see Table~\ref{tab:pr2} for the qubit positions in the frequency combs used for Fig.~3(b-c).
\begin{table}[ht]
\begin{center}
    \begin{tabular}{|p{25mm}|p{25mm}|p{25mm}|p{25mm}|p{25mm}|p{25mm}|}
    \hline
    $\Delta\omega/(2\pi)$, MHz & $\omega_2/(2\pi)$, GHz & $\omega_3/(2\pi)$, GHz & $\omega_7/(2\pi)$, GHz & $\omega_5/(2\pi)$, GHz & $\omega_6/(2\pi)$, GHz\\ \hline\hline
    \textbf{35} & 5.8045 & 5.8395 & 5.8745 & 5.9095 & 5.9445 \\ \hline
    \textbf{40} & 5.7945 & 5.8345 & 5.8745 & 5.9145 & 5.9545 \\ \hline
    \textbf{45} & 5.7845 & 5.8295 & 5.8745 & 5.9195 & 5.9645 \\ \hline
    \textbf{50} & 5.7745 & 5.8245 & 5.8745 & 5.9245 & 5.9745 \\ \hline
    \end{tabular}
    \caption{\label{tab:pr2}The qubit frequencies $\omega_k$ extracted from transmission measurements Fig.~3(a) using the fit function Eq.~4 for the fixed qubit parameters from Table~\ref{tab:pr1}.}
\end{center}
\end{table}

\section*{Pulse power calibration}
\begin{figure}[t]
\centering
\includegraphics[width=0.6\columnwidth]{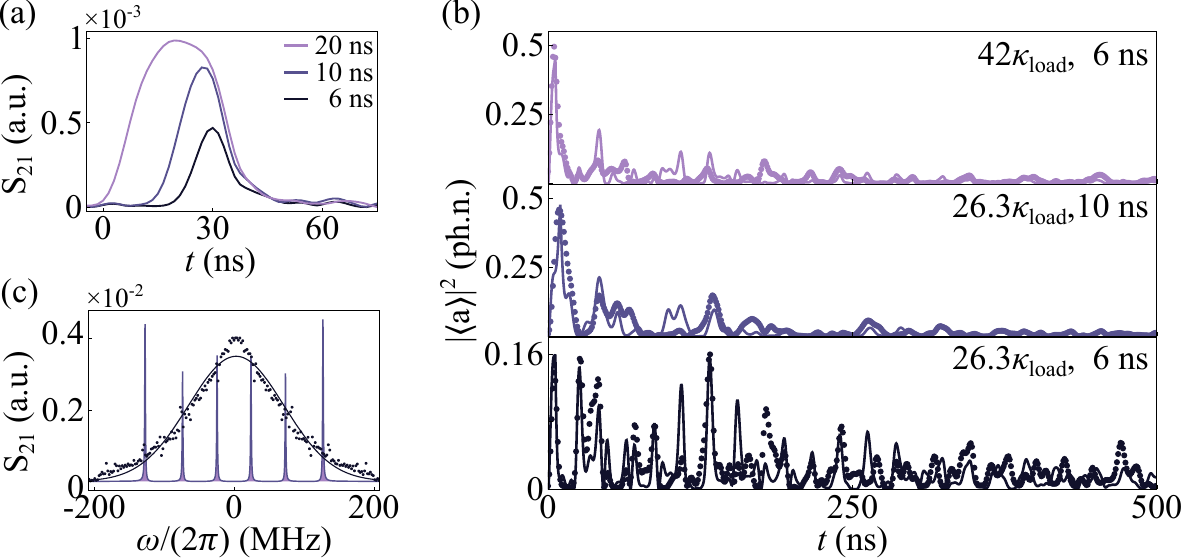}
\caption{(a)) Pulses amplitude measured directly after upconversion as the function of time. (b) Squared amplitude of the transmitted signal as a function of time $t$ measured close to the revival saturation point. The frequency detuning between qubits $\Delta\omega/(2\pi) = 50$\,MHz. (c) Fourier transform of the measured 6 ns pulse (black points) with Gaussian fit (black solid line) and $\Delta\omega/(2\pi) = 50$\,MHz frequency comb transmission spectrum (purple solid line) as the function of frequency, $2\sigma=80.4$\,MHz. Filled pink area shows the spectral overlap between the excitation pulse and the frequency comb.} \label{fig:pulse_4_pulse}
\end{figure}

For the time domain measurements, the system is excited by the short ($6$\,ns) pulse which is generated using upconversion of the arbitrary waveform generator (AWG) output signal, see Fig. \ref{Sfig:pulse_2}(a). Since the time-resolution of the AWG is limited to $2$\,ns and the upconversion mixer has the bandwidth of 500\,MHz, such a short pulse reaches only a reduced amplitude, as shown in Fig.~\ref{fig:pulse_4_pulse}(a). 

To estimate the drive amplitude $\eta$ in Eq. 3 for the given output power of the upconversion chain, we measure the pulse revival dynamic close to the breaking point where revivals are distorted with the $6$ and $10$\,ns pulses, as well as at $4$\,dB higher input power. We find that for $26.3\eta$ drive amplitude for the breaking point pulse and $42\eta$ drive amplitude for the $4$\,dB higher pulse, all three measurements can be fitted with the same scaling parameter $B=383271$, as shown in Fig.~\ref{fig:pulse_4_pulse}(b). The same scaling factor was used to fit all the consecutive measurements with the corresponding change of $\eta$ for lower power measurements.

%%%%%%%%%%%%%%%%%%%%%%%%%%%%%%%%%%%%%%%%%%%%%%%%%%%%%%%%%%%
%%%%%%%%%%%%%%%%%%%%%%%%%%%%%%%%%%%%%%%%%%%%%%%%%%%%%%%%%%%%

The estimated drive amplitude characterizes the input power. However, the input is not mode-matched with the frequency comb, thus, the number of photons in the system is lower as the frequency comb acts like a filter. In Fig.~\ref{fig:pulse_4_pulse}(c), we show the Fourier transform of the $6$\,ns pulse measured directly after the upconversion. The pulse is fitted with the Gaussian  function (solid black line)
\begin{equation}\label{gaus}
f(x) = A e^{-\frac{(\omega-\mu)^2}{2\sigma^2}},
\end{equation}
where $A$ is a dimensionless amplitude, $\mu\approx2$\,MHz is an expected value, and $2\sigma=80.4$\,MHz is the bandwidth. Ideally, the excitation pulse should match the full bandwidth of the frequency comb, shown with the purple solid line in Fig.~\ref{fig:pulse_4_pulse}(c), which for the frequency detunings $g\lesssim \Delta\omega \lesssim 2g$ is around 200\,MHz. The overlap between the pulse and frequency comb is demonstrated with the filled pink area and equal to $0.04$ of the full pulse power.

\end{document}